\documentclass{JHEP}
\usepackage{amsmath,amssymb}
\newcommand{\pd}{\partial}
\newcommand{\tr}{\mathop{\rm tr}}

\newcommand{\spec}{\mathop{\rm Spec}}
\newcommand{\rank}{\mathop{\rm rank}}
\newcommand{\corank}{\mathop{\rm corank}}
\newcommand{\A}{{\mathcal A}}
\newcommand{\hh}{{\mathcal H}}

\newcommand{\I}{{\mathbb I}}
\newcommand{\F}{{\mathcal F}}
\newcommand{\G}{{\mathfrak G}}
\renewcommand{\L}{{\mathfrak L}}
\title{M[any] Vacua of IIB\thanks{Work supported by RFBR
grant \# 99-01-00190, INTAS grant \# 950681, and Scientific School
support grant 96-15-0628}}%
\author{Corneliu Sochichiu\\ Institutul de Fizic\u a Aplicat\u a
A\c S, str. Academiei, nr. 5, Chi\c sin\u au, MD2028 MOLDOVA\\
and\\ \thanks{Present address} Bogoliubov Laboratory of
Theoretical Physics, Joint Institute for Nuclear Research, 141980
Dubna, Moscow Reg., RUSSIA
\\
E-mail: \email{sochichi@thsun1.jinr.ru}}%
\preprint{\hepth{0004062}}
\abstract{Description of the spectrum of fluctuations around a
commutative vacuum solution, as well as around a solution with
degenerate commutator in IIB matrix model is given in terms of
supersymmetric Yang--Mills (YM) model. We construct explicitly the
map from Hermitian matrices to YM fields and study the dependence
of the spectrum and respective YM model on the symmetries of the
solution. The gauge algebra of the YM model is shown to contain
local reparameterisation algebra as well as Virasoro one.}
\begin{document}
\section*{Introduction}

IKKT, or IIB matrix model \cite{ikkt}, is a statistical mechanic
model which was introduced as a nonperturbative regularisation of
IIB superstring model in the Schild gauge \cite{Schild:1977vq}. It
is the string analog of lattice gauge models and provides a tool
for non-perturbative numerical study of the string theory
(M-theory) \cite{Ambjorn:2000bf}--\cite{Nishimura:2000ds}.

The picture one has in the IKKT model is closely related to the
Connes' approach to the noncommutati\-ve geometry
\cite{Connes:1996gi}. In this approach a manifold is described in
terms of the so called Connes' triple $(\A,\Delta,\hh )$, rather
than as a set of points. In this triple, $\A$ is the algebra of
bounded operators (algebra of functions), $\Delta$ is some
elliptic differential operator (Laplace or Dirac operator, for
simplicity we consider Euclidean signature) and $\hh$ is the
Hilbert space where $\A$ and $\Delta$ are represented. ``Points''
of such a manifold can be identified with the spectra of some
position operators built from the triple, other local and global
characteristics can as well be extracted from it.

In the case of IKKT matrix model the r\^{o}le of $\A$ is played by
Hermitian matrices with ``smooth'' eigenvalue distribution and
bounded square trace in the limit $N\rightarrow\infty$, while the
$\Delta$ and $\hh$ are represented by the background solution and
the adjoint representation of $U(N)$.

Compactifications of this model to $d$-dimensional noncommutative
tori was shown to result in noncommutative Yang--Mills (YM) models
in respective dimensions \cite{Connes:1998cr}. In particular,
compactification to a circle yields the
Banks--Fischler--Shenker--Susskind (BFSS) matrix model which was
introduced as a non-perturbative regularisation of the light-cone
membrane action in $D=11$, \cite{Banks:1997vh}.

The compactification in Ref. \cite{Connes:1998cr} is introduced as
a restriction of the background solution to satisfy some
periodicity conditions modulo gauge transformations. The
respective periods are identified with the periods of the torus on
which the model is compactified.

From the other hand, given a BPS background, i.e. a solution to
equations of motion which preserves a part of the supersymmetry,
one can map the space of $N\times N$ Hermitian matrices of IKKT
model to the space of real or matrix valued functions on some
non-commutative manifold. Under certain conditions this map is
isomorphism of algebras where the matrix product is mapped into
noncommutative or star product of functions due to the
noncommutativity of the manifold. The properties of the manifold
are exclusively determined by the respective BPS background
solution.

Explicitly the BPS solution is given by a set of matrices having
scalar commutators. In the case when the matrix of commutators is
nondegenerate (or for the subset on which it is nondegenerate) one
can easily construct the map from the space of arbitrary Hermitian
matrix fluctuations around respective BPS background to the space
of functions on a noncommutative manifold \cite{Ambjorn:1999ts}.
The limit of commutative manifold corresponds to infinite
commutator of the solution.

The case when the commutator is degenerate corresponds just to an
opposite situation. One may argue, and this was the main reason
for arriving at actual paper, that only commutative solutions give
the global minimum of the IKKT actions. However, as we show in the
Section 2., this is not exactly the case, since due to homogeneity
of the IKKT action in bosonic fields, the respective action
vanishes on purely bosonic solutions. Unfortunately, this
regretful error committed in the pioneering work \cite{ikkt}, was
reproduced in the succeeding papers\footnote{See e.g. most recent
paper \cite{Ishibashi:2000eu}.}. In spite of this disappointing
discovery the study of the spectrum of fluctuations around a
commutative solution still presents some interest, in particular
due to the fact that, as it will be seen below, the degeneracy in
the commutator of the solution corresponds to increasing the
dimensionality of corresponding noncommutative YM model, which
gives a ``physically'' different model (compare this with the
situation with the first and second class constraints,
\cite{Dirac:1950eu}).

To obtain a noncommutative $U(1)$ YM model from the matrix
fluctuations one has to impose some irreducibility condition,
while allowing certain degeneracy to the background solution one
comes to description of a ``vector fibre bundle'' over the
noncommutative manifold which leads to ``nonabelian''
noncommutative YM. Respective degeneracy can be interpreted as
compactification on many coinciding branes, while the
nondegenerate case corresponds to a single brane \cite{ikkt}.

The objective of the actual paper is to analyse the relation
between the background commutative/degenerate vacuum solution and
the properties of the resulting YM model. Although, the
commutative solution is a particular case of a generic BPS
solution, as we already mentioned, it is ``physically'' different,
since it corresponds to a singular limit of the generic case.

The plan of the paper is as follows. In the next section we give a
brief account of IIB matrix model. After that we consider a BPS
solution, show that it corresponds to a vanishing action and
consider in more details the commutative case. We impose a set of
conditions such a solution must respect, and build explicitly the map
between the matrix fluctuations and functions of noncommutative
manifold which appears to be a ``noncommutative'' product of two
commutative manifolds dual to each other. This allow to find in the
Section 4. a representation of IIB matrix model in terms of YM fields
for both commutative and generic degenerate case.

Finally, we discuss the results and consider the consequences and
possible generalisations of the actual analysis.

\section{The IIB Matrix Model}

The IKKT, or IIB matrix model, is described by the classical
action:
\begin{equation}\label{cl_action}
  S=-\frac{1}{g^2}\tr\left(
  \frac{1}{4}[A_\mu,A_\nu]^2+\frac{1}{2}\bar{\psi}\Gamma^\mu[A_\mu,\psi]\right),
\end{equation}
where $A_\mu$, $\mu=1\dots D=10$, $\psi$ and $\bar{\psi}$ are
$N\times N$, $N\rightarrow \infty$ Hermitian matrices which act on
$N$ dimensional Hilbert space $\hh$, with scalar product
$\eta^{\dag}\eta$, $\eta\in\hh$. Spinor matrices $\bar{\psi}$ and
$\psi$ also carry the $SO(10)$ Majorana--Weyl spinor index, while
bosonic ones $A_\mu$ carry vector one. We assume summation
convention for repeated indices.

Equations of motion corresponding to the action (\ref{cl_action})
look as follows
\begin{align}\label{eqm1}
  [A_\mu,[A_\mu,A_\nu]]-[\bar{\psi},\Gamma_\mu \psi]&=0, \\ \label{eqm2}
  \Gamma^\mu[A_\mu,\psi]=[\bar{\psi},A_\mu]\Gamma^\mu&=0.
\end{align}

This model possesses a $\mathcal{N}=2$ supersymmetry corresponding
to the transformations,
\begin{align}\label{susy1}
  \delta_{(1)} A^\mu&=i\bar{\epsilon}\Gamma^\mu \psi, \\
  \delta_{(1)} \psi&=\frac{i}{2}[A_\mu,A_\nu]\Gamma^{\mu\nu}\epsilon,
\end{align}
and,
\begin{align}\label{susy2}
  \delta_{(2)} A_\mu=0, \\
  \delta_{(2)} \psi= \xi,
\end{align}
where $\epsilon$ and $\xi$ are Majorana--Weyl spinors, as well as
a SU(N) gauge symmetry,
\begin{align}\label{gauge}
  A_\nu \rightarrow U^{-1}A_\nu U, \\
  \psi \rightarrow U^{-1}\psi U, \\
  \bar{\psi} \rightarrow U^{-1}\bar{\psi} U,
\end{align}
and  $U$ is $N\times N$ unitary matrix. Since only the adjoint
representation is in use this symmetry is $SU(N)/\mathbb{Z}_N$
rather that $U(N)$. This will not remain true if ``matter'' fields
are included, in this case $U(1)$ and $\mathbb{Z}_N$ will act
nontrivially on the fields in the fundamental representation of
$U(N)$

Another symmetry is already mentioned $SO(10)$ ``covariance'' of
10-dimensional vectors and spinors.

\section{The Vacuum}
In the limit $N\rightarrow \infty$ equations
(\ref{eqm1},\ref{eqm2}) have an important class of purely bosonic
($\psi=0$) solutions which preserve a part of the supersymmetry
(BPS solutions). These solutions are given by $A_\mu=p_\mu$, where
matrices $p_\mu$ satisfy,
\begin{equation}\label{bps}
  [p_\mu,p_\nu]=iB_{\mu\nu},
\end{equation}
with $B_{\mu\nu}$ as $u(N)$ matrix proportional to the unity one,
$B_{\mu\nu}\sim \I$. For finite $N$ the commutator (\ref{bps})
cannot be satisfied but only mod some matrix vanishing in the
sense of operator norm as $N\rightarrow \infty$. A solution of the
described type can be realised e.g. by shift-and-clock operators
on functions defined on a lattice
\cite{Connes:1998cr,Ambjorn:1999ts}.

Naively, action (\ref{cl_action}) computed on the solution
(\ref{bps}) should be equal to \cite{ikkt},
\begin{equation}\label{bps_action}
  S_{BPS}=\frac{N}{4g^2}(B_{\mu\nu})^2.
\end{equation}

As one can see this expression does not realise the minimum (even
local one) unless $B_{\mu\nu}=0$, since one can devaluate the
action (\ref{bps_action}) by a smooth variation of the solution,
e.g. $\delta p_\mu=\lambda p_\mu$, which decreases the action if
$\lambda<0$. The problem is accomplished by the fact that the
variation of the action is \emph{first order} in $\lambda$, which
should not be the case if one varies a background solution.
Indeed, this problem is solved if we see that for any finite $N$
the action computed on a purely bosonic solution of equations of
motion vanishes. Thus, for any finite $N$ the bosonic part of the
action (\ref{cl_action}) can be rewritten as,
\begin{equation}\label{total_zero}
  S=\frac{1}{4g^2}\tr A_\nu [A_\mu,[A_\mu,A_\nu]],
\end{equation}
which vanishes identically if $A_\mu$ is a solution to
(\ref{eqm1}) with zero $\psi$.

The case with infinite $N$, however, is still indefinite since it
depends how the limit $N\rightarrow\infty$ is achieved. One can
obtain the limit either from $p_\mu(N)$ satisfying equations of
motion or from ones not satisfying them, in the last case the
value of action depends on how fast $p_\mu(N)$ approaches a
solution as $N$ goes to infinity. Thus, for a configuration
$p_\mu(N)$, satisfying,
\begin{equation}\label{ie}
  [p_\mu,p_\nu]=iB_{\mu\nu}\I(N),
\end{equation}
where $\I(N)=\I+\epsilon(N)$ is a unity matrix approaching
sequence. In order to approach (\ref{bps}), one has to require,
\begin{align}\label{tr_eps}
  \tr \epsilon(N)=-N \\ \label{eps->0}
  \lim_{N\rightarrow\infty}\epsilon (N)=0,
\end{align}
where the limit is computed using the operator norm,
$\|\epsilon\|=\sup_{\eta^{\dag}\eta=1}\sqrt{\eta^{\dag}
\epsilon^{\dag}\epsilon\eta}$, $\eta\in\hh$. Eqs. (\ref{tr_eps})
and (\ref{eps->0}) are contradictory unless we restrict the space
of vectors $\eta$ on which the operator norm is computed e.g. by
vectors having a fixed finite number of nonzero coordinates in the
limit $N\rightarrow\infty$. In this case, the action
(\ref{cl_action}) computed on on this sequence of configurations
reads,
\begin{equation}\label{sn}
  S_{BPS}(N)=\frac{1}{4g^2}B^2\tr\I(N)^2=\frac{1}{4g^2}B^2(\tr
  \epsilon(N)^2-N),
\end{equation}
where the factor $(\tr\epsilon(N)^2-N)$ is ambiguous. The
additional requirement that for all $N$ the configuration $p_\mu$
to be a solution to (\ref{eqm1}) eliminates the ambiguity since
for any $N$ one has $S_{BPS}(N)=0$, so we assume this requirement
to be satisfied.

The vanishing of the classical BPS action results in ``equality''
in the factor $e^{-S}$ between solutions $p_\mu$ with different
commutators (\ref{bps}), either BPS or non-BPS. The BPS vacua are,
however, preferred since one expects to have no loop corrections
to them \cite{ikkt}.

The above problems are absent in the case of commutative
solutions, i.e. when $B_{\mu\nu}=0$. Before considering this case
in more details, consider a generic solution (\ref{bps}). By a
proper linear transformation $B_{\mu\nu}$ can be brought to the
``standard'' form having block diagonal form with two
dimensional blocks of the form,
\begin{equation}\nonumber
  \begin{pmatrix}
    0 & -1 \\
    1 & 0
  \end{pmatrix}
,
\end{equation}
and a $r\times r$  zero block, corresponding to zero modes of
$B_{\mu\nu}$.

The set of matrices $(p_\mu)$ is split in this case in three
subsets $(p_i,q^i,p_a)$, where $p_i$ and $q^i$ form canonical
conjugate pairs, and $p_a=z_a^\mu p_\mu$, $a=1,\dots,r=\corank B$,
correspond to zero modes $z_a^\mu$ of $B_{\mu\nu}$,
$B_{\mu\nu}z_a^\nu=0$,
\begin{align}\label{split_conj}
  &[p_i,q^j]=-i \delta_i^j,\\ \label{split_0}
  &[p,p]=[q,q]=0.
\end{align}

Intuitively, at this stage one can see that $p_\mu$ corresponding
to a nondegenerate part of $B_{\mu\nu}$ carries $1$ (one-particle
Hamiltonian) degree of freedom while one corresponding to zero
mode caries $2$ degrees. The second degree of freedom should be
given by the canonical conjugate $q^a$, if it exists. We hope this
statement will become more transparent below.

For a nonzero $B_{\mu\nu}$, as well as for zero one it can be seen
that the set of adjoint operators $P_\mu=[p_\mu,\cdot]$, (which
act on $u(N)$ matrices), is a commutative one,
\begin{equation}\label{pcommute}
  (P_\mu P_\nu-P_\nu
  P_\mu)a=[p_\mu,[p_\nu,a]]-[p_\nu,[p_\mu,a]]=i[B_{\mu\nu},a]=0.
\end{equation}
This fact common with one that $P_\mu$ are selfadjoint with
respect to the scalar product,
\begin{equation}\label{scprod}
  (a,b)=\tr a^\dag b,
\end{equation}
where $a^\dag$ stands for the hermitian conjugate matrix, means
that $P_\mu$ can be diagonalised even for noncommutative set of
matrices $p_\mu$ satisfying eq. (\ref{bps}). One can, therefore,
decompose any hermitian $N\times N$ matrix in the orthogonal basis
of $P_\mu$ eigenmatrices. Having this decomposition at hand we
will construct in the next section the map from the space of
Hermitian matrices to the space of real functions on some
noncommutative manifold.

In the case when $B_{\mu\nu}=0$ one can diagonalise not only
adjoint operators $P_\mu$ but also all matrices $p_\mu$.

We say that $p_\mu$ is a nondegenerate commutative vacuum
background solution if the following are satisfied:
\begin{itemize}
  \item All $p_\mu$ are commutative,
\begin{equation}\label{comm}
  [p_\mu,p_\nu]=0.
\end{equation}
  \item Matrices $p_\mu$ are (functionally) independent, this
  implies also linear independence,
\begin{equation}\label{lin}
  \alpha^\mu p_\mu=0\Rightarrow\text{all }\alpha^\mu=0.
\end{equation}
  \item They satisfy,
  \begin{align}\label{trace1}
    \tr p_\mu&=0, \\ \label{trace2}
    \tr p_\mu p_\nu&=0, \qquad \text{for }\mu\neq\nu.
  \end{align}
  \item Eigenvalues of $p_\mu$ form a $D$-dimensional irregular lattice
  with sites symmetrically distributed with respect to the origin (the
  centre of the lattice), in the range
  $|\lambda |\leq \Lambda$\footnote{One can alternatively chose eigenvalues
  to lie in other compact domain, e.g. a box,
  $\Lambda_\mu\leq\lambda_\mu\leq\Lambda_\mu$.}, where
  $|\lambda|=\sqrt{\lambda_\mu^2}$. In the limit
  $N\rightarrow\infty$ the lattice becomes dense.

  This implies that one can rearrange the matrix labels (which is
  equivalent to a gauge transformation) in such a way that
  eigenvalues $\lambda_\mu$ are distributed according to ``smooth''
  and ``monotonic'' functions of the matrix index $s_\nu$,
  $\lambda_\mu(s_\nu)$, which is the $s_\nu$-th eigenvalue, where
  $s_\nu$ form a linear $D$-dimensional lattice,
  \begin{align}\label{eig_prop1}
    \det \left(\frac{\pd \lambda_\mu (s)}{\pd s_\nu}\right)>0,\qquad &
    \\ \label{eig2}
    \lambda_\mu (-s)=-\lambda_\mu (s),\qquad &\\ \label{eig3}
    |\lambda_\mu(s+1)-\lambda_\mu(s)|=
    \left|\frac{\pd \lambda_\mu (s)}{\pd s_\nu}\right|=o(N^{-1}),
    \qquad &N\rightarrow \infty, \Lambda=\text{fixed}.
  \end{align}
\end{itemize}
The last gives the smooth distribution of eigenvalues in the limit
$N\rightarrow\infty$. In this limit quantity $\Lambda$ plays the
role of UV cutoff. This implies that for a finite but large values
of $N$ one is justified to manipulate with $\lambda_\mu (s)$ as
with quasicontinuous quantities.

The authentic continuum limit is achieved by sending $\Lambda$ to
infinity after the limit $N\rightarrow\infty$ is reached.

The eigenvalue problem in the case of commuting $p_\mu$ is
equivalent to the Cartan decomposition problem in Lie algebra
$u(N)$, see e.g. \cite{Hermann:1994eu}. In this context the above
enlisted conditions say that $p_\mu$ must form the orthogonal
basis in a $D$-dimensional plane of Cartan subalgebra of $su(N)$
to which \emph{no} root is orthogonal.

\section{The Map}

As usual any matrix $a$ is a function of a pair of labels
$(s,s')$, $a=a(s,s')$. Matrix product and trace are given by,
respectively,
\begin{align}\label{sprod}
  (a\cdot b)(s',s'')&=\sum_{\{s_\mu\}}a(s',s)b(s,s''),\\ \label{strace}
  \tr a&=\sum_{\{s_\mu\}}a(s,s).
\end{align}

In what follows we will use also quasicontinuous notations where
the sums we will substitute by the integrals. Taking into account
that the increment of $s$ in (\ref{sprod}) and (\ref{strace}) is
$d s=\Delta s=1$ these sums can be written as following integrals,
\begin{align}\label{cprod}
  (a\cdot b)(s',s'')&=\int ds\,a(s',s)b(s,s''), \\
  \label{ctrace}
  \tr a&=\int ds\, a(s,s).
\end{align}
Where the range of integration is given by $-\frac{N_\mu}{2}\leq
s_\mu \leq \frac{N_\mu}{2}$ and $\prod_{\mu}N_\mu =N$.

The unity matrix is given by the $\delta$ symbol,
\begin{equation}\label{unity}
  \I(s,s')=\delta_{ss'}\equiv \delta(s-s').
\end{equation}
Note, that in these notations $\delta (0)=1$, this differs from
the usual ``continuous'' $\delta$-function by a diverging factor
of order $N$.

The limit $N\rightarrow \infty$ is achieved in the following way.
First, one introduce an UV cutoff $L$ and rescale $s\rightarrow
\frac{2L}{N}s$. Thus the increment $\Delta s$ becomes really small
and one has the genuine integration in eqs.
(\ref{sprod}-\ref{ctrace}). The cutoff removing is obtained in the
limit $L\rightarrow \infty$. Since we plan to identify $p_\mu$
with momentum operators the continuous $\lambda_\mu$ spectrum
corresponds to noncompact manifolds, while to have the compact
result one has to keep the discreteness of the spectrum and sums
instead of the integrals.

In this picture matrices $p_\mu$ look as follows,
\begin{equation}\label{ps}
  p_\mu (s,s')=\lambda_\mu (s)\delta (s-s'),
\end{equation}
where $\lambda_\mu(s)$ scan the eigenvalue lattice of $p_\mu$,
while an arbitrary diagonal matrix $\varpi$ looks like,
\begin{equation}\label{pi}
  \varpi (s,s')=\varpi(s)\delta (s-s').
\end{equation}

It is not difficult to compute the action of operator $P_\mu$ on
an arbitrary matrix $a(s,s')$,
\begin{multline}\label{Pact}
  (P_\mu a)(s,s')=[p_\mu,a](s,s')=\\
  \int ds'' \, (\lambda_\mu
  (s)\delta(s-s'')a(s'',s')-a(s,s'')\lambda_\mu(s'')\delta(s''-s'))\\
  =(\lambda_\mu(s)-\lambda_\mu(s'))a(s,s').
\end{multline}

One immediately finds that the spectrum of $P_\mu$ is given by
eigenvalues $k_\mu =\lambda_\mu(t)-\lambda_\mu(t')$ for any $t$
and $t'$ taking values in the lattice $\{s_\mu\}$, and corresponding
to eigenvectors,
\begin{align}\label{eigs}
  f_{tt'}(s,s')&=\delta (s-t)\delta (s'-t'),\\
  P_\mu
  f_{tt'}&=\left(\lambda_\mu(t)-\lambda_\mu(t')\right)f_{tt'}.
\end{align}

For large values of $N$ eigenvalues $k$ become degenerate, the
respective eigenfunctions maybe labelled by eigenvalue $k_\mu$ and
a label $t$ counting the degenerate states,
\begin{equation}\label{kdeg}
  f_t(k)=\int dt' \delta (t'-s(\lambda(s)-k))f_{tt'}=
  f_{ts(\lambda(s)-k)},
\end{equation}
where the function $s(\lambda)$ is the inverse to the
$\lambda(s)$: $s(\lambda(s))=s$, $\lambda(s(\lambda))=\lambda$.

Now, consider the matrix $q^\mu$ defined by,
\begin{equation}\label{q}
  q^\mu=-i\frac{\pd}{\pd k_\mu}\left.\int dt f_t
  (k)\right|_{k=0}=
  iA_\alpha{}^\mu(s)\delta'{}^\alpha (s'-s),
\end{equation}
where,
\begin{equation}\label{A}
  A_\alpha{}^\mu(s)=\left.\frac{\pd s_\alpha}{\pd
  \lambda_\mu}\right|_{\lambda=\lambda(s)}=\left(\left(\frac{\pd \lambda}{\pd
  s}\right)^{-1}(s)\right){}_\alpha{}^\mu,
\end{equation}
and $\delta'{}^\alpha (s)=\frac{\pd}{\pd s_\alpha}\delta (s)$.

As one can immediately see the commutator of $p_\nu$ with $q^\mu$
is the canonical one,
\begin{equation}\label{pq}
  [p_\mu,q^\nu]=-i\delta_\mu^\nu \I,
\end{equation}
i.e. $p_\mu$ and $q^\nu$ form a canonical conjugate pair and can
be interpreted as, respectively, momentum and coordinate
operators, as we anticipated earlier. Strictly speaking, such
pairs of operators exist only in the limit $N\rightarrow \infty$,
i.e. when the spectrum of either $q$ or $p$ is continuous.

From $q$'s one can construct a matrix $E(k)=e^{ik_\mu q^\mu}$.
Matrix $E(k)$ is nondegenerate and it is an eigenvector of $P_\mu$
corresponding to the value $k$,
\begin{equation}\label{eq}
  [p_\mu,E(k)]= k_\mu E(k),
\end{equation}
having the squared norm,
\begin{equation}\label{normE}
  \|E(k)\|^2=\tr E^{\dag}(k)E(k)=N.
\end{equation}

For a finite $N$ the dimensionality of $k$-eigenspace depends on
$k$ and decrease as $k$ approaches to $2\Lambda$ since matrix
edges are approaching for large values of $k$. When $N\rightarrow
\infty$ this difference disappears and one can regard the spaces
with different $k$ as isomorphic. This approximation is accurate
as soon as $|t-t'|\ll N$, where $\lambda(t)-\lambda(t')=k$. Matrix
$E(k)$ can be considered as the eigenvalue shift operator,
\begin{equation}\label{ek}
  E(k)f(k')\sim f(k+k').
\end{equation}
Due to isomorphism and nondegeneracy of $E(k)$, eigenvectors
corresponding to a nonzero $k$ are given by the product of zero
vectors and matrix $E(k)$ as follows,
\begin{equation}\label{ft}
  E_t(k)=f_t(0)E(k),
\end{equation}
where $f_t(0)$ is the basis in the space of diagonal matrices
which form the zero space. It is transparent that this basis is
orthogonal in the space of $N\times N$ if $f_t(0)$ form an
orthonormal basis in the space of diagonal matrices, the norm of
$E_t(k)$ coinciding with that of $f_t(0)$. Thus, given an
orthonormal basis in the zero space of $P_\mu$ one can spread it
out through all $k$-eigenspaces using eq. (\ref{ft}).

Since eigenvectors of $P_\mu$ form a complete orthonormal set in
the space of $N\times N$ matrices, any Hermitian matrix $a$ can be
expanded in the basis of $P_\mu$ eigenvectors,
\begin{equation}\label{eig}
  a=\sum_{t,k}\tilde{a}_t(k)f_t(k)=\sum_{t,k}\tilde{a}_t(k)
  f_t(0)e^{ik\cdot q}.
\end{equation}

Orthonormality of the spectrum assures the inverse transformation,
\begin{equation}\label{inv_eig}
  \tilde{a}_t(k)=\frac{1}{\|f_t(0)\|^2}\tr e^{-ikq}f_t^{*}(0)a.
\end{equation}

From the properties a vacuum solution one can trace that arbitrary
matrix commuting with $p_\mu$ is a diagonal matrix (\ref{pi}), and
it may be represented as a function of $p_\mu$,
\begin{equation}
  \varpi=\varpi (p)\equiv \varpi(s(p)),
\end{equation}
where $\varpi (s(\lambda))|_{\lambda=p}$ is the same function as
in (\ref{pi}), but computed on diagonal matrices $p_\mu$.

One may expand $\varpi$ in multiple ways e.g. in Fourier series as
function of $p_\mu$,
\begin{equation}
  \varpi=\sum_{z}\tilde{\varpi}(z)e^{ip\cdot z},
\end{equation}
where $z_\mu$ are points of the lattice of eigenvalues of operator
$Q^\mu=[q^\mu,\cdot]$,
\begin{equation}
  [q^\mu,e^{ipz}]=-z^\mu e^{ipz}.
\end{equation}

Since operators $Q^\mu$ are also commutative selfadjoint operators
(they commute also with $P_\mu$) their spectrum satisfy,
\begin{equation}
  \tr e^{ip(z-z')}=N \delta_{zz'},
\end{equation}
which gives the ground for inverse transformation,
\begin{equation}\label{invpi}
  \tilde{\varpi} (z)=\frac{1}{N}\tr e^{-ipz} \varpi .
\end{equation}

Using the above decomposition for the diagonal matrices one can
rewrite the expansion eq. (\ref{eig}) for arbitrary Hermitian
matrix $a$ as follows,
\begin{align}\label{mfour}
  &a=\sum_{z,k}\tilde{a}(z,k)e^{-ikz}e^{ipz}e^{ikq}=
  \sum_{z,k}\tilde{a}(z,k)e^{ipz+ikq},\\ \label{mfour_inv}
  &\tilde{a}(k,z)=\frac{1}{N}\tr a e^{-ipz-ikq},\\ \label{mfour_herm}
  &\tilde{a}^*(k,z)=\tilde{a}(-k,-z).
\end{align}

At this point one can identify the space of Hermitian matrices
with the space of real functions on the ``noncommutative
manifold'' $\spec(P) \times \spec (Q)$, where $\spec (P)$ and
$\spec (Q)$ are the varieties of eigenvalues of $P$, and
respectively of $Q$. In fact when $N\rightarrow \infty$ with
$\Lambda$ fixed $\spec (P)$ form a compact manifold, while $\spec
(Q)$ tends to a noncompact lattice. Continuum limit for this
lattice is achieved in the limit $\Lambda\rightarrow\infty$.

This map is given by\footnote{We use the same character to denote
both the matrix and corresponding function the difference being
that for functions we write explicitly its arguments, i. e. if $a$
denote some matrix we  write $a(z,l)$ for the corresponding
function and viceversa.},
\begin{equation}\label{m->f}
  a\mapsto a(x,l)=\sum_{z,k}\tilde{a}(z,k)e^{ilz+ikx},
\end{equation}
where $\tilde{a}(z,k)$ is defined by eq. (\ref{mfour_inv}).

This map can be turned backward, therefore, it is one-to-one. The
corresponding matrix is given by,
\begin{equation}\label{f->m}
  a=\sum_{k,z}\tilde{a}(k,z)e^{ipz+ikq},
\end{equation}
now, $\tilde{a}(z,k)$ is the Fourier transform of the function
$a(x,l)$,
\begin{equation}\label{four2}
  \tilde{a}(z,k)=\frac{1}{N}\sum_{x,l}a(x,l)e^{-ilz-ikx}.
\end{equation}

Under the correspondence (\ref{m->f}---\ref{four2}) the matrix
product is mapped into the star product given by,
\begin{equation}\label{prod->*}
  (a\cdot b)\mapsto a\star b(x,l)=\left.e^{-\frac{i}{2}
  \left(\frac{\pd^2}{\pd x' \pd l}-\frac{\pd^2}{\pd x\pd l'}\right)}
  a(x,l)b(x',l')\right|_{x'=x \atop l'=l},
\end{equation}
while the trace corresponds to the lattice integration over $x$
and $l$,
\begin{equation}\label{tr<->int}
  \tr a \rightarrow \sum_{l\in \spec (P) \atop x\in \spec (Q)}
  a(x,l)\equiv \int dxdl\, a(x,l).
\end{equation}

Commutators with $p_\mu$ and with $q^\mu$ give the lattice analogs
of differentiation with respect to $x^{\mu}$ and respectively
$l_\mu$,
\begin{align}\label{pd1}
  [p_{\mu},a]\rightarrow -i\frac{\pd}{\pd x^\mu}a(x,l), \\
  [q^\mu,a]\rightarrow -i\frac{\pd}{\pd l_\mu}a(x,l).
\end{align}

\section{The Spectrum}
Let us now return back to the Matrix Model action
(\ref{cl_action}) and consider arbitrary fluctuations around the
vacuum solution $p_\mu$ satisfying (\ref{comm}---\ref{trace2}),
\begin{equation}\label{fluc}
  A_\mu=p_\mu+ga_\mu,
\end{equation}
here $a_\mu$ is an arbitrary hermitian matrix.

Perturbed action looks as follows,
\begin{equation}\label{a_action}
  S=-\tr\left(
  \frac{1}{4}([p_\mu,a_{\nu}]-[p_\nu,a_\mu]+g[a_\mu,a_\nu])^2
  +\bar{\psi}\Gamma^\mu[(p_\mu+ga_\mu),\psi]\right).
\end{equation}

Now, let use the correspondence between matrices and functions to
map the fluctuation $a_\mu$ to function $a_\mu(x,l)$. Then, action
(\ref{a_action}) is rewritten as follows,
\begin{equation}\label{s_func}
  S=-\int d^Dx\, d^Dl
  \left(\frac{1}{4}\F_{\mu\nu}^2(x,l)+\bar{\psi}\star\Gamma^\mu\nabla_\mu
  \psi(x,l)\right),
\end{equation}
where $d^Dx$ and $d^Dl$ are invariant measures on $\spec(Q)$ and
$\spec(P)$ respectively. They are given by eigenvalue distribution
densities for $P$ and $Q$. Also,
\begin{align}\label{F}
  &\F_{\mu\nu}=i\frac{\pd }{\pd x^\mu}a_\nu(x,l)-
  i\frac{\pd }{\pd x^\nu}a_\mu(x,l)
  -g[a_\mu,a_\nu]_{\star}(x,l) \\ \label{nabla}
  &\nabla_\mu\psi(x,l)=i\frac{\pd}{\pd x^\mu}\psi(x,l)
  -g[a_\mu,\psi]_{\star}(x,l),
\end{align}
where $[\cdot,\cdot]_{\star}$ stands for the star commutator
defined as,
\begin{equation}\label{scom}
  [a,b]_{\star}=a\star b-b\star a.
\end{equation}

Action (\ref{s_func}) give the \emph{exact} description of IIB
matrix model in $N\rightarrow\infty$ limit in terms of functions
on the manifold $\spec(P)\times\spec(Q)$. This action possesses a
huge gauge symmetry given by the following transformations by a
star-unitary function $U(x,l)$,
\begin{align}\label{gau_a}
  a_\mu(x,l)&\rightarrow U^{-1}\star a_\mu \star U(x,l)-
  \frac{i}{g}U^{-1}\star\frac{\pd}{\pd x^\mu}U(x,l)\\ \label{gau_psi}
  \psi(x,l)&\rightarrow U^{-1}\star \psi (x,l),
\end{align}
where $U^{-1}(x,l)=U^{\star}(x,l)$, and star conjugate function is
given by the function corresponding to the Hermitian conjugate
matrix and in this particular case coincides with the complex
conjugate function,
\begin{equation}
  U^{\star}(x,l)=(U^{\dag})(x,l)=U^{*}(x,l).
\end{equation}

Global gauge group $G$ can be identified with ``constant''
transformations $U$, satisfying $\pd_\mu U=0$. This means that $U$
may depend only on the momentum parameter $l$, i.e. $U=U(l)$. At
first sight it seems that the gauge group is $U(1)$ group
localised along $\spec(P)$, since gauge symmetry is realised by
scalar functions and not matrix valued ones. Indeed, the
``global'' group is Abelian,
\begin{equation}\label{com_gauge}
  U(l)\star U'(l)=U(l)U'(l)=U'(l)\star U(l),
\end{equation}
but this commutativity does not hold for local transformations.

Consider the algebra of infinitesimal gauge transformations $\G$.
This is algebra of real functions with star commutator,
\begin{equation}\label{gauge-alg}
  [f,g]_{\star}=i h,\qquad f,g,h \in \G.
\end{equation}

Natural basis for this algebra form following functions,
\begin{equation}\label{Ls}
  L_{k,z}(x,l)=e^{ilz+ikx}.
\end{equation}
The commutator for generators $L_{k,z}$ looks as follows,
\begin{equation}\label{LL}
  [L_{k,z},L_{k',z'}]_{\star}=2i\sin \frac{1}{2}(k'z-k z')L_{k+k',z+z'}.
\end{equation}

Thus, one may interpret the model (\ref{s_func}) as an ordinary
(commutative) Yang--Mills model with the local algebra of gauge
transformations (\ref{LL}).

It is worthwhile to note that although the model we got is defined
in the terms the flat space, the gauge symmetry the model has
contains the group of local $x$-reparameterisations as a subgroup,
whose infinitesimal transformations are generated by,
$L_{\epsilon}=\epsilon^\mu (x)l_\mu$,
\begin{equation}\label{rep}
  [L_{\epsilon},f(x)]=\epsilon^\mu(x)\pd_\mu f(x),
\end{equation}
but this is not all. It is not difficult to show that algebra
(\ref{LL}) contains Virasoro subalgebra which may serve as an
indication that action (\ref{s_func}) also describe the string
spectrum. Indeed, consider a functions $\theta(x,l)$ defined mod
$2\pi$ and its canonical conjugate $w(x,l)$,
\begin{equation}\label{theta_pi}
  [w,\theta]_{\star}=-i.
\end{equation}
Then, generators $L_n=e^{in\theta}w$ satisfy,
\begin{align}\label{vir1}
  &L_n^{\star}=L_{-n} \\ \label{vir2}
  &[L_n,L_m]_{\star}=i(n-m)L_{n+m},
\end{align}
which is exactly the classical Virasoro algebra.

The above results can be readily extended to the case of a generic
background commutator $B_{\mu\nu}$. In this case we assume the
subset $(p_i,p_a)$ to satisfy conditions
(\ref{comm})---(\ref{eig3}). Operators $q^i$, and respective
eigenfunctions are already known, one has only to find the
counterparts to $p_a$. Repeating the derivations of the previous
section for this particular case, one comes to the map from
matrices to functions $f(x^\mu,l^a)$, with star product defined
as,
\begin{equation}\label{star2}
  (a\star b)(x,l)=e^{-\frac{i}{2}\tilde{C}^{\mu\nu}\frac{\pd^2}
  {\pd x^\mu \pd x^{\prime\nu}}
  -\frac{i}{2}\left(\frac{\pd^2}{\pd x^{\prime a}\pd l_a}
  -\frac{\pd^2}{\pd x^a \pd l'_a}\right)}a(x,l)b(x',l'),
\end{equation}
tensor $\tilde{C}^{\mu\nu}$ is defined by,
\begin{equation}\label{wavy_C}
  \tilde{C}^{\mu\nu}B_{\nu\alpha}=B_{\alpha\nu}\tilde{C}^{\nu\mu}=
  \Pi^\mu_\alpha,
\end{equation}
where $\Pi^\mu_\alpha$ is the projector to the space orthogonal to
zero modes of $B_{\mu\nu}$.

The YM action (\ref{s_func}) keeps in this case the same form
except the integration is done over $d^Dx\,d^rl$, and the star
product is given by eq. (\ref{star2}). As we see, zero modes of
$B_{\mu\nu}$ lead to extra integrations in (\ref{s_func}) over
$dl$, or extra gauge (and, respectively, physical) degrees of
freedom.

\section{Generalisations and Conclusions}

In analysing the spectrum of fluctuations around a vacuum solution
of the classical action (\ref{cl_action}) we made several
assumptions about the respective solutions. Namely, we required
$p_\mu$ to form an independent and nondegenerate commutative set
of matrices. This lead us to the action (\ref{s_func}) for fields
on the product of the manifold corresponding to the spectrum of
coordinate operator and its dual given by the spectrum of momentum
operator. This representation contains a broad symmetry including
invariance with respect to local reparameterisations as well as
$2D$ conformal symmetry, and can be interpreted as a Yang Mills
model with gauge algebra given by (\ref{LL}).

The above assumptions seem natural, but one may pose a question:
what may happen if one gives up some of them?

Consider first the case when all conditions are respected except
the number of independent matrices is not equal to $D$ but is
smaller. Let, in particular, matrices $p_\mu$ be expressed as
linear combinations of the independent subset $(p_\alpha)$
$\alpha=1,\dots,p+1<D$,
\begin{equation}\label{lind}
  p_\mu=\xi_\mu^\alpha p_\alpha,
\end{equation}
where $\rank \|\xi_\mu^\alpha\|=p+1$.

In this case, one can find the position operators corresponding to
independent $p_\alpha$, and a generic hermitian matrix will be
expandable in the basis of the matrices $e^{ip_\alpha
z^\alpha+ik_\alpha q^\alpha}$. Therefore, after manipulations like
in previous section, hermitian matrices are now represented by the
real functions defined on the spectrum of the subset of
\emph{independent} $p_\alpha$ and $q^\alpha$,
$\alpha=1,\dots,p+1$.

As a result, one has action (\ref{a_action}) mapped into a
$(p+p)$-dimensional action which corresponds to the reduction of
the action (\ref{s_func}) of $(D+D)$-dimensional YM model to a
$(p+p+2)$ dimensional plane given by $\xi_\mu^\alpha$.

One may describe the above situation as a localisation of IIB
matrix model to a $p$-brane in contrast to $(D-1)$-brane we had
initially in (\ref{s_func}).

Consider now the opposite case, i.e. when all matrices $p_\mu$
together fail to possess nondegenerate set of eigenvalues. The
last means that i) there exist eigenvalue sets $(\lambda_\mu)$
whose eigenspace are degenerate and, therefore, ii) there are
matrices $\pi $ commuting with all $p_\mu$ but not being functions
of $p_\mu$. It is not difficult to show that the set of these
matrices generate at most central extended Lie algebra $\L$,
\begin{equation}\label{cl}
  [\pi_a,\pi_b]=ic_{ab}\cdot\I+f^c_{ab}\pi_c.
\end{equation}

Thus, one is able map the generic Hermitian matrix $a$ from $u(N)$
algebra to a $\L$-valued function $a(x,l)$. In particular, when
$\L=u(n)$ this corresponds to a ``nonabelian'' generalisation of
the action (\ref{s_func}). The ``physical'' interpretation of this
solution is the localisation of IIB matrix model to $n$ copies of
coinciding branes, \cite{ikkt}.

All above says that the spectrum of the matrix model is in a
strong dependence of the vacuum solution chosen. Due to the limit
$N\rightarrow\infty$, the different background solutions $p_\mu$
lead to different continuum models (\ref{s_func}). One may
conjecture that the respective models may be classified by the
rank $r$ of $B_{\mu\nu}$ and intersection of stabiliser groups of
each $p_\mu$, or by the symmetry of the vacuum.

In this paper we considered the solution with the smallest
possible group of symmetry giving the $D$-dimensional action. In
fact, this configuration has the ``largest'' measure (entropy
factor), or better to say ``the moduli space'' among the
commutative solutions. The opposite extreme is given by the
``solution'' where all $p_\mu$ are proportional to unity matrix
$p_\mu=\lambda_\mu \I$ which describes the $u(N)$ YM model
localised on a point, i.e. the original matrix model. The ``moduli
space'' of such solution is parameterised by just $D$ numbers
$\lambda_1,\dots,\lambda_D$.

Thus, in IIB matrix model one has a plenty of vacua. For a finite
$N$ these vacua are connected through the fluctuations
corresponding to zero modes of the Hessian matrix
$S_{\mu\nu}\equiv\frac{\pd^2 S}{\pd A_\mu \pd A_\nu}(p)$. As
$N\rightarrow\infty$, we expect such fluctuations to fall out of
the class of allowed functions, namely continuous $L^2$-integrable
functions, and the vacua to become separated.

\acknowledgments
I am grateful to P.~Pyatov, A.~Nersesian, and T.~Bakeev for useful
discussions and critical remarks.
\providecommand{\href}[2]{#2}\begingroup\raggedright\endgroup

\end{document}